\documentclass[12pt]{iopart}
\usepackage{graphicx}

\begin{document}
\input epsf
\title{The antiferromagnetic transition of UPd$_2$Al$_3$ break-junctions:
A new realization of $N$-shaped current-voltage characteristics}

\author{Yu.~G.~Naidyuk \footnote[3]{To whom correspondence should be
addressed (naidyuk@ilt.kharkov.ua)}, K.~Gloos \dag, I.~K.~Yanson,
N.~K.~Sato \ddag  }

\address{B.~Verkin Institute for Low Temperature Physics and
Engineering, National Academy of Sciences of Ukraine, 61103,
Kharkiv, Ukraine}

\address{\dag  Nano-Science Center, Niels Bohr Institute fAFG,
Universitetsparken 5, DK-2100 Copenhagen, Denmark}

\address{\ddag Department of Physics, Graduate School of Science, Nagoya
University, Nagoya 464-8602, Japan}

\date{\today}



\begin{abstract}
We have investigated metallic break junctions of the heavy-fermion
compound UPd$_2$Al$_3$ at low temperatures between 0.1\,K and 9\,K
and in magnetic fields up to 8\,T. Both the current-voltage $I(V)$
characteristics and the d$V$/d$I(V)$ spectra clearly showed the
superconducting ($T_{\rm c}\simeq 1.8$\,K) as well as the
antiferromagnetic ($T_{\rm N}\simeq $14\,K) transition at low
temperatures when the bias voltage is raised. The junctions with
lateral size of order 200\,nm had huge critical current densities
around $5\times 10^{10}\,$A/m$^2$ at the antiferromagnetic
transition and {\em hysteretic} $I(V)$ characteristics. Degrading
the quality of the contacts by {\em in situ} increasing the local
residual resistivity reduced the hysteresis. We show that those
hysteretic $I(V)$ curves can be reproduced theoretically by
assuming the constriction to be in the thermal regime. It turns
out that these point contacts represent non-linear devices with
$N$-shaped $I(V)$ characteristics that have a negative
differential resistance like an Esaki tunnel diode.
 \pacs{73.63.-b, 74.25.Fy, 74.50.+r}

\end{abstract}

 \maketitle

\section{Introduction}
Point-contact (PC) spectroscopy is widely used to study the
interaction of conduction electrons with elementary excitations or
quasiparticles in conducting solids \cite{Yanson,Duif}.
Energy-resolved PC spectroscopy is possible when the inelastic
relaxation length of electrons in the contact region $\Lambda =
{\rm min} \{l_{\rm in},\sqrt{l_{\rm el} l_{\rm in}/3}\}$ (here
$l_{\rm el}$ and $l_{\rm in}$ are the elastic and the inelastic
mean free path of the electrons) is larger than the size or
diameter $d$ of the contact. In the opposite case of $\Lambda \ll
d$ and a short phonon scattering length $l_{\rm ph} < d$, the
excess electron energy dissipates in the constriction. This Joule
heating increases the temperature inside the contact when a bias
voltage is applied \cite{Kohlrausch,Verkin,Kulik}.

Therefore the interpretation of the PC data requires to find out
the regime of charge transport. $l_{\rm el}$ does not depend on
energy and can be determined rather accurately for the PC region.
$l_{\rm in}$ depends on energy, and no method exists to calculate
it reliably. To identify the transport regime becomes especially
important for PCs with complex systems like heavy-fermion,
high-T$_c$, or Kondo-lattice compounds that typically have large
electrical resistivities because of their strong electron
correlations.

A.~Wexler \cite{Wexler} derived \begin{equation} \label{Rwex} R(T)
=  \frac {16 \rho l}{3\pi d^2} + \beta\,\frac{\rho (T)}{d}
\end{equation} for the PC resistance $R$ as function of
temperature $T$ and contact size $d$. The parameter $\beta\simeq
1$ varies slowly as function of $l_{\rm el}/d$, and $\beta = 1$
for large contacts $d \gg l_{\rm el}$. Wexler's formula
interpolates between the ballistic Sharvin ($l = l_{\rm el} \gg
d$) and the diffusive Maxwell ($l\ll d$) resistance. The latter
describes transport as in the bulk material.

Since Sharvin's resistance does not depend on temperature,
differentiating Eq.~(\ref{Rwex}) with respect to temperature
yields \begin{equation} \label{d-T} d= \frac{d\rho/dT}{dR/dT}
\end{equation} for the size of the contact. This is considerably
more reliable for deriving $d$ than Eq.~(\ref{Rwex}) itself. The
main reason is that the residual resistivity in the PC region can
strongly differ from the bulk $\rho_0$, for example due to the
stress exerted while forming the contact. Eq.~(\ref{d-T}) was
experimentally verified for PCs with simple metals by Akimenko
\etal \cite{Akimenko}.

The same method can be applied to heavy-fermion compounds. They
show at low temperatures power-law dependencies of their
electrical resistivities $\rho(T)  = \rho_0 + AT^n$ ($n = 1, 2, 3$
for the various compounds investigated) which was  also revealed
in the PC resistances \cite{Gloos96}. Heavy-fermion compounds
typically have large $A$ coefficients because of their strong
electron correlations, which makes it straightforward to measure
$dR/dT$ just above $T_c$. For those high-resistivity
superconducting (SC) metals the local normal-state residual
resistivity in the PC region \begin{equation} \label{rho_0} \rho_0
= d~ \delta R \label{dR} \end{equation} can be extracted from the
drop $\delta R$ of the contact resistance due to SC. Such a
relationship has been found for a number of heavy-fermion SCs over
a wide range of contact sizes \cite{Gloos97,Naidyuk1,Gloos98}.

\begin{figure}
\centering
\includegraphics[width=8cm,angle=0]{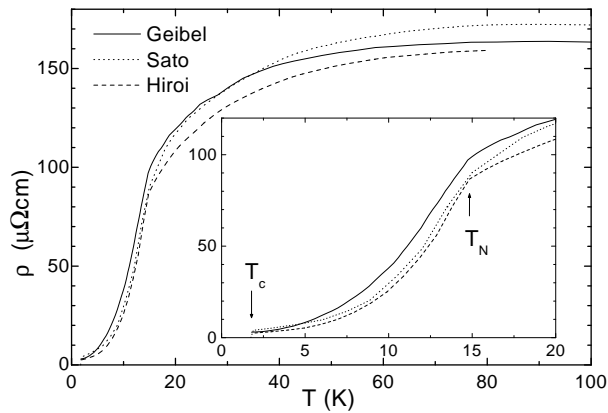}
\caption[]{Electrical resistivity $\rho (T)$ of polycrystalline
UPd$_2$Al$_3$ (solid curve \cite{Geibel})  and of two single
crystals along the basal ab-plane (dotted curve \cite{Sato,Sato1},
dashed curve \cite{Hiroi}). The inset shows $\rho (T)$ at low
temperatures. Arrows mark the SC and the AFM transition,
respectively.} \label{rhot}
\end{figure}

Here we present experiments on PCs between two pieces of the
heavy-fermion compound UPd$_2$Al$_3$ \cite{Geibel}, using the
technique of mechanically-controllable break junctions. Compared
to the conventional spear-anvil type technique to form point
contacts, break junctions have much better mechanical stability.
But more importantly breaking the sample at low temperatures in
the ultra-high vacuum region of the refrigerator avoids oxidation
of the freshly broken surfaces of the contact interface.
UPd$_2$Al$_3$ becomes antiferromagnetic (AFM) at $T_N
\simeq$14\,K. It is SC below $T_c \simeq 1.8\,$K. We have observed
huge non-linearities of the PC resistances and even hysteretic
$I(V)$ characteristics. We derived the contact size and the
residual resistivity in the PC region according to
Eqs.~(\ref{d-T}) and (\ref{rho_0}), respectively. We found that
the very short elastic mean-free path in the constriction $l_{\rm
el}\ll d$ points to at least the diffusive regime of electron
transport through the PC. Considering also the small inelastic
mean-free path in UPd$_2$Al$_3$, reflected by the steep $\rho(T)$
rise with temperature around the AFM transition in
Fig.~\ref{rhot}, we applied the thermal model developed in
Refs.~\cite{Verkin,Kulik} for the case $l_{\rm el}, l_{\rm in}\ll
d$ to take into account the locally increased temperature in the
PC when a bias voltage is applied. Using the experimental $\rho
(T)$ in Fig.~\ref{rhot}, this model described quite well the
observed $I(V)$ characteristics and their modification with
temperature, reproducing also the hysteretic features.

\section{Experiment}
We have investigated three UPd$_2$Al$_3$ single crystals. Two of
them had one long side in the c-direction of the hexagonal crystal
lattice, one sample had it in the perpendicular ab-direction. A
$0.5 - 0.7\,$mm deep notch was cut into the middle of the $\sim
1\times1\times5\,$ mm$^3$ large UPd$_2$Al$_3$ bars using a diamond
saw. This defined the break position. Each sample was glued
electrically isolated onto a flexible metal bending beam. Twisted
pairs of voltage and current leads were attached with silver epoxy
to both sides of the sample, which was then  mounted onto the
mixing chamber inside the vacuum can of the dilution refrigerator.
The temperature could be varied between 0.1 and 9\,K. With a
micrometer screw the bending beam is bent at low temperatures,
breaking the sample at the notch. The resistance of the break
junction, that is  its lateral contact size, could be adjusted
mechanically with the micrometer screw. For further details of the
experimental setup see Refs.~\cite{Naidyuk1,Gloos98}.

\begin{figure}
\centering
\includegraphics[width=8cm,angle=0]{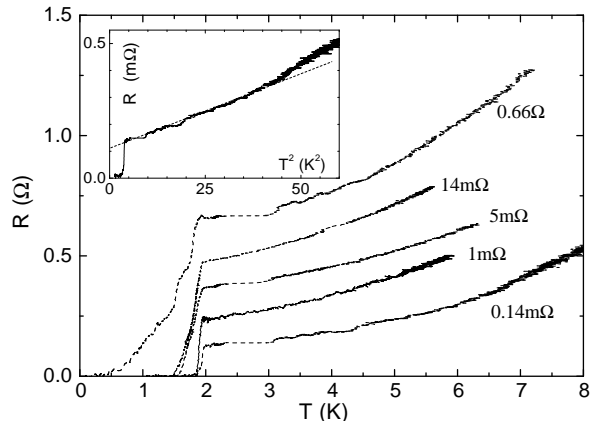}
\caption[]{Resistance $R$ of UPd$_2$Al$_3$ break junctions along
the ab-direction versus temperature $T$ before breaking (bottom
curve) and with increasing PC resistance. The curves, except the
upper one, are scaled along the $R$-axis to fit into the same
window. The resistance $R_n$ at $T\geq T_c$ just above the SC
transition is indicated for each curve. The contact of the upper
curve is about 200\,nm wide (see text). Dashed horizontal lines
indicate missing data between 2.2 and 3\,K in some of the $R(T)$
curves. This was due to an instability of the mixing chamber of
the refrigerator while slowly warming up. The inset shows $R(T)$
vs $T^2$ of the unbroken sample. The straight dotted line
describes the contribution of the $A$ coefficient. The current
excitation was $I = 1\,$mA for the contacts with $R_n \le
5\,$m$\Omega$, 0.5\,mA for the 14\,m$\Omega$, and 2.5\,$\mu$A for
the 0.66\,$\Omega$ contact. It was chosen small enough to not
degrade $R(T)$.} \label{rt} \end{figure}

At room temperature the resistance of the samples with the notch
was about 5\,m$\Omega$, corresponding to the approximate
geometrical cross-section and a contact size of 0.2\,mm. Note that
the notch only defines the macroscopic position of the junction,
the microscopic contact is less well defined. After removing the
sample from the refrigerator, the surface of the junction was not
mirror-like or smooth as expected for a single crystal. The
fracture was usually tilted with respect to the direction of the
notch, and thus the crystal axis. Therefore the current flow
through  the contact might deviate slightly from the direction
defined by the long side of the sample. Magnetic fields up to 8\,T
could be applied perpendicular to the bending mean, that is
perpendicular to the long side of the samples and to the ideal
direction of current flow.

The $I(V)$ characteristic and the differential resistance
$dV/dI(V)$ were recorded by injecting a DC current $I$ superposed
by a small alternating current $\delta I$ and measuring the
differential voltage drop $V$. Its alternating part $\delta V$ was
detected using the standard lock-in technique.

\section{Results}
All three UPd$_2$Al$_3$ single crystals showed qualitatively the
same results. Therefore we concentrate here on one of them, that
with the long side in ab-direction. Figure \ref{rt} shows the
temperature dependence of the resistance $R(T)$ of the break
junctions below 9\,K before breaking and of several contacts after
successively reducing the contact size by increasing the bending
force. The superconducting transition at 1.8\,K as well as the
$\sim T^2$ increase above $T_c$ like in the bulk samples is
clearly seen. Occasionally $R(T)$ changes in small steps. The
reason for this is that UPd$_2$Al$_3$ single crystals are quite
brittle. They also have a large thermal expansion with respect to
the bending beam above $\sim 1\,$K. When the temperature changes,
the stress in the contact region changes. This stress is sometimes
partly released, slightly varying the contact size or the local
residual resistivity and, thus, $R(T)$.

With increasing PC resistance the SC transition broadens. We
believe that this is mainly due to the stress in the PC area when
the sample is broken and the contact being formed. Additional
broadening is caused by the extremely small critical supercurrent
which suppresses the Sharvin resistance at low temperatures and
small excitation voltages, see also the discussion below. On
increasing the temperature the critical current decreases, so that
Sharvin's resistance is again added to the total resistance.
However, its contribution to $R(T)$ is small since for the
investigated contacts Sharvin's resistance is much smaller than
Maxwell's resistance. Changing the force on the bending beam
changes the contact size and the stress there, too. This allows
us, although in an uncontrolled manner, to vary {\em in situ} the
local resistivity at the PC.

\begin{figure}
\centering
\includegraphics[width=8cm,angle=0]{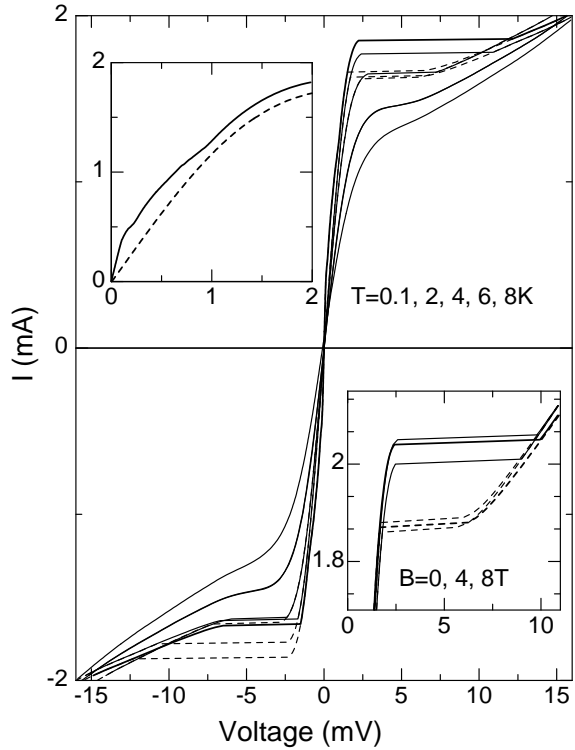}
\caption[]{$I(V)$ characteristics of the UPd$_2$Al$_3$ break
junction with $R_n = 0.66\,\Omega$ at the indicated temperatures.
Solid (dashed) lines correspond to sweeps with increasing
(decreasing) current. The hysteretic loops become smaller when the
temperature rises, vanishing above $\sim 5\,$K. The upper inset
shows $I(V)$ below (0.1\,K, solid line) and above (2\,K, dashed
line) the SC transition in extended scale. The lower inset shows
part of the $I(V)$ curves at $T$=0.1\,K and the indicated magnetic
fields.} \label{iv1} \end{figure}

\begin{figure}
\centering
\includegraphics[width=8cm,angle=0]{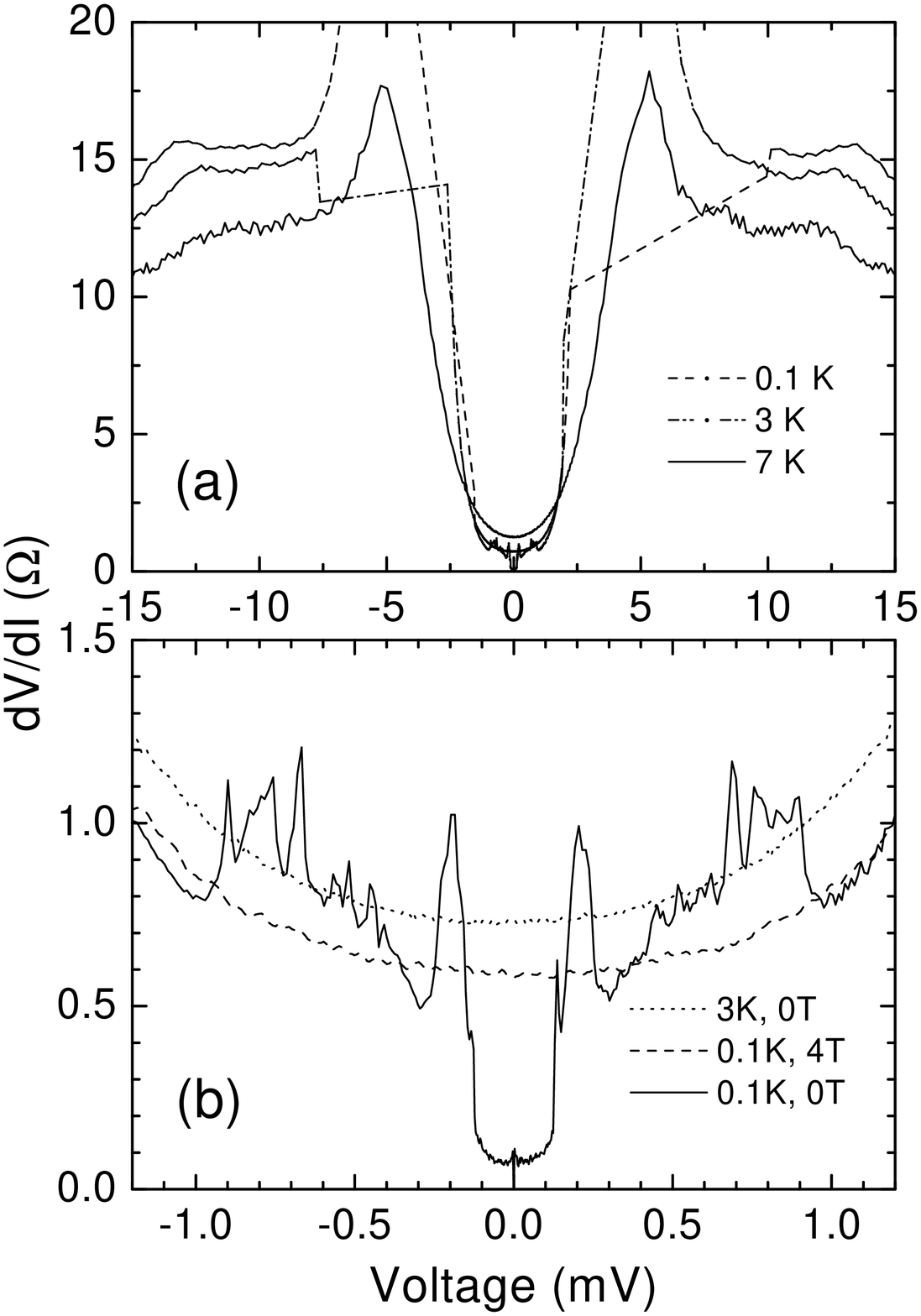}
\caption[]{(a) Differential resistance $dV/dI$ of the
UPd$_2$Al$_3$ break junction from Fig.\,\ref{iv1} at 0.1, 3, and
7\,K  at high biases. Below $\sim 5\,$K some of the curves are
discontinuous around $\pm$5\,mV, indicated by the dashed lines.
(b) $dV/dI$ of the same break junction at low biases. The SC
anomaly has disappeared at 3\,K or at $B$=4\,T, that is well above
either $T_c\simeq$1.8\,K or $B_c\simeq$3.5\,T of UPd$_2$Al$_3$. }
\label{dv1} \end{figure}

Figure \ref{iv1} shows for the UPd$_2$Al$_3$ break junction with
$R_n = 0.66\,\Omega$ as example how the $I(V)$ curves typically
change with temperature. At low temperatures $I(V)$ is strongly
hysteretic. At higher temperatures the hysteresis smears out and
transforms into an inflection point that corresponds to the
pronounced $dV/dI$ maxima above about 5\,K in Fig.~\ref{dv1}(a).
Large magnetic fields up to $B = 8\,$T only slightly modified the
$I(V)$ curves at 0.1\,K by reducing the size of the hysteretic
loop. A 4\,T field as well as the temperature above $T_c$
completely suppressed the superconducting features, a zero-bias
minimum of the differential resistance accompanied by a series of
spikes, see Fig.~\ref{dv1}(b).

\begin{figure*}[]
\begin{center} \epsfxsize=16cm \epsfbox{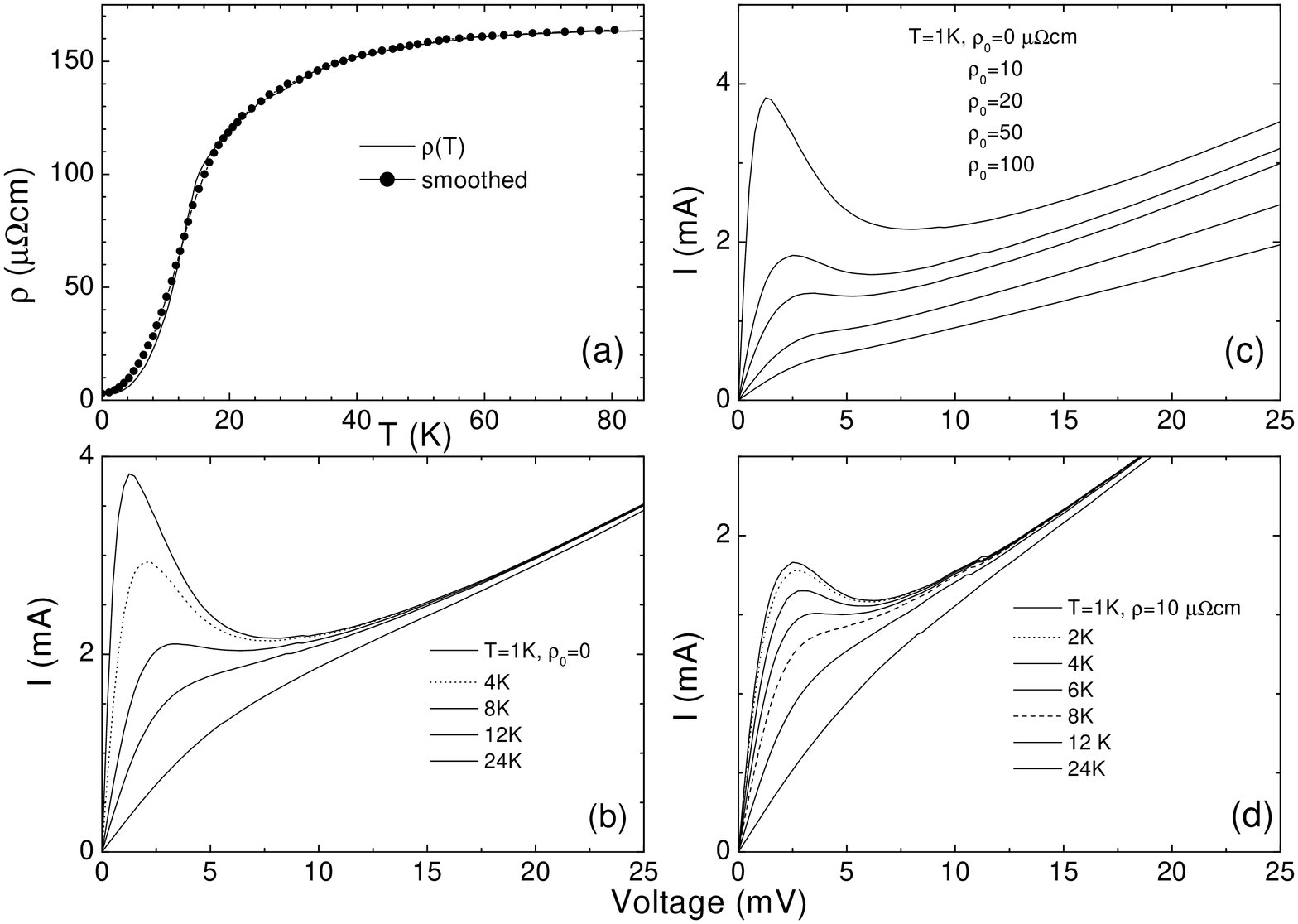}
\caption{(a) Smoothed $\rho (T)$ (symbols) used for modelling the
PC. The solid line shows Geibel`s $\rho (T)$ from Fig.\,\ref{rhot}
for comparison. (b) $I(V)$ characteristics of the UPd$_2$Al$_3$ PC
at different temperatures, calculated according to
Eq.\,(\ref{IVT}) for $d$=200 nm and assuming $\rho_0 = 0$. (c)
Modification of the calculated $I(V)$ at 1\,K by adding the
residual resistivity $\rho_0$ to $\rho (T)$. (d) Calculated $I(V)$
curves  at different temperatures for $\rho_0=10\mu\Omega$cm and
$d$=200\,nm. } \label{ivcg} \end{center} \end{figure*}

\section{Discussion}

We start the analysis by deriving the size $d$ of the contacts.
Above $T_c$ both the specific resistivity and the contact
resistance vary with the same $A\,T^2$ power law. According to
Eq.~(\ref{d-T}) the contact size $d = A_{bulk}/A_{PC}$. Literature
values for the $A_{bulk}$ coefficient range from 0.15 to
$0.25\,\mu\Omega$cm\,K$^{-2}$, for example in
Refs.~\cite{Geibel,Sato1,Hiroi}. In part this variation could be
due to micro-cracks in the bulk samples which spoils the
geometrical factor. Therefore we choose the average $A_{bulk} =
0.20\,\mu\Omega$cm\,K$^{-2}$, which coincides with that in
\cite{Sato1}. The absolute error in $d$ can then amount up to
about $\pm 33\,\%$, but the relative accuracy needed to compare
the different contacts is much better. In this way the contact in
Fig.~\ref{iv1} has $d \approx 200\,$nm.

We can now directly read off the critical current density from the
$I(V)$  data in Fig.~\ref{iv1}. For the AFM transition the current
density reaches up to $5\times 10^{10}\,$Am$^{-2}$. At the SC
transition, marked by the $dV/dI$ maximum in Fig.~\ref{dv1}(b),
the critical current density approaches $1.5\times
10^{10}\,$Am$^{-2}$. Both values are {\em lower} bounds for the
{\em kinetic} critical current densities because they include
local heating of the PC discussed below.

According to Eq.~(\ref{rho_0}) the $\delta R = 0.66\,\Omega$
resistance drop due to SC results then in a normal-state residual
resistivity $\rho_0 \approx 13\,\mu\Omega$cm. This is about three
times larger than the bulk $\rho_0$, estimated from $R(T)$ of the
unbroken junction in Fig.~\ref{rt}.

This $\delta R$ includes a possible contribution from the
Josephson effect: The differential resistance vanishes completely
within a very narrow ($\sim 10\,\mu$V) voltage range around zero
bias, barely seen in Fig.~\ref{dv1}. The much broader ($\sim
0.3\,$mV) zero-bias minimum has a plateau of around
$0.10\,\Omega$, fitting well the ballistic Sharvin resistance
calculated using the known contact diameter. This agreement
supports our interpretation that we are dealing not with multiply
connected contacts but with single contacts. Taking into account
Sharvin's resistance would slightly reduce the calculated local
residual resistivity from $13\,\mu\Omega$cm to $11\,\mu\Omega$cm.

The elastic electron mean free path at low temperatures can be
estimated using the typical metallic $\rho l \simeq 2.5\times
10^{-15}\,\Omega$m$^2$ (here $l$ is the elastic mean free path and
$\rho$ and $l$ values taken from \cite{Geibel}) as $l_{\rm el}
\approx 20\,$nm. This leads to the inequality $l_{\rm el} \simeq
20\,$nm$\ll 200\,$nm$\simeq d$ for two of the important length
scales of the constriction, implying that these PCs are at least
in the diffusive regime. However,  heavy-fermion compounds
typically have a large residual resistivity and/or already at low
temperatures a strongly increasing electrical resistivity, they
are very likely in the thermal regime \cite{Naidyuk}.

The ballistic Sharvin resistance is then negligible, and the PC
resistance can be described by Maxwell's resistance
\begin{equation} R(T) \simeq \rho(T)/d~. \label{Rm} \end{equation}
In contrast to a ballistic contact, where energy dissipates far
away form the contact region, now all energy is released in the
constriction. This increases its temperature with bias voltage.
Assuming the Wiedemann-Franz law to be valid, the temperature in
the center of the PC  depends on the applied voltage via
\cite{Kohlrausch,Verkin} \begin{equation} \label{T-V} T^2 =
T^2_{\rm bulk} + \frac {V^2}{4L}, \end{equation} When the
temperature $T_{\rm bulk}$ of the bulk sample vanishes, the
contact temperature varies linearly with bias voltage like
$T=V/2\sqrt{L}$. Using the standard Lorenz number $L = L_0 =
2.45\cdot 10^{-8}\,$V$^2$K$^{-2}$, a 1\,mV bias voltage will raise
the temperature of the contact by 3.2\,K.

In the thermal regime the $I(V)$ characteristic of the contact
depends on the temperature-dependent electrical resistivity $\rho
(T)$ according to \cite{Verkin,Kulik} \begin{equation} \label{IVT}
I(V) = Vd \int_0^1  \frac{{\rm d}x}{\rho(T \sqrt{1-x^2}~)}
\end{equation} where $T$ is defined by Eq.~(\ref{T-V}). We used
the smooth curve in Fig.~\ref{ivcg}(a) to approximate the
experimental $\rho(T)$, but omitted the SC transition.

The calculated $I(V)$ curves in Fig.~\ref{ivcg}(b) had maxima at
around $2 - 3\,$mV, which results in a hysteresis for up- and
downward sweeps when the junction is driven by a current source.
These maxima are the larger the steeper the drop in $\rho (T)$
around $T_N$. They decrease and become broader with increasing
residual resistivity, see Fig.~\ref{ivcg}(c).

With voltage biasing we would expect to recover the full $I(V)$
chracteristics without hysteresis. However, this would require to
install small resistors near the sample in parallel and in series
with the break junction to cut off its bistability, see for
example Ref.~\cite{Leadbeater89}.  This was not practical in our
experiments because in each cool down we wanted to investigate
many break junctions over a wide range of resistances.

\begin{figure}
\centering
\includegraphics[width=8cm,angle=0]{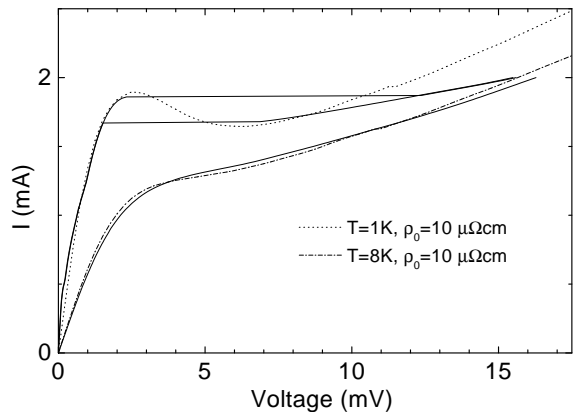} %
\vspace{0cm} \caption[]{Comparison between the experimental $I(V)$
characteristics (solid curves) from Fig.~\ref{iv1} compared to the
calculated ones (dashed curves) from Fig.~\ref{ivcg}(d) at low and
at high temperatures. The bottom calculated curve is multiplied by
0.9 along the $I$ axis.} \label{ivcexp}
\end{figure}

Figure \ref{ivcexp} shows that the theoretical $I(V)$ describe
well the experimental data, including the width of the hysteretic
features, using $d = 200\,$nm and $\rho_0 = 10\,\mu\Omega$cm.
These are the only two adjustable parameters. And they fit well
those parameters that have been derived above independently from
the measured $R(T)$. This agreement strongly supports our
interpretation that local (at the PC) thermal effects determine
the behavior of our UPd$_2$Al$_3$ break junctions.

Surprisingly, the standard Lorenz number $L_0$ yields the best fit
to the experimental $I(V)$ curves while in bulk UPd$_2$Al$_3$
$L(T)$ rises from 0.6$L_0$ below 1\,K up to $\approx 15\,L_0$ at
24\,K because of the dominant heat transport by phonons
\cite{Hiroi}. This implies that at the PC the phonon channel is
closed, and heat is carried away by electrons only.

\begin{figure}
\centering
\includegraphics[width=8cm,angle=0]{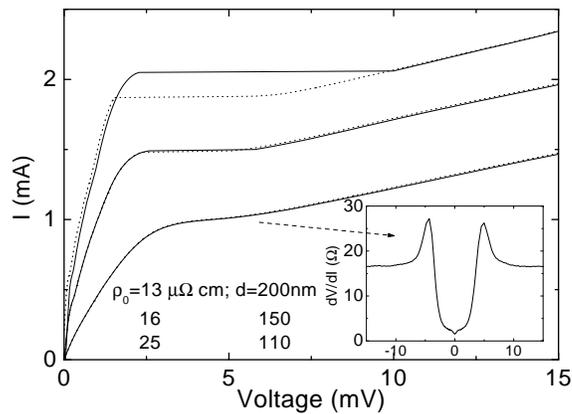}
\caption[]{$I(V)$ characteristics of different UPd$_2$Al$_3$ break
junctions at $T\simeq$0.1\,K. The current was swept upward (solid
lines) and downward (dotted lines). The residual resistivity
$\rho_0$ and the contact size $d$, derived from $R(T)$, are given
for each contact. Inset shows $dV/dI$ for the bottom $I(V)$ curve.
} \label{ivc1} \end{figure}

With increasing residual resistivity the hysteresis of the
experimental $I(V)$ curves in Fig.~\ref{ivc1} transforms into an
inflection point, corresponding to the maxima in $dV/dI$ (inset of
Fig.~\ref{ivc1}). This trend agrees with the theoretical curves in
Fig.~\ref{ivcg}(c). However, it seems that with increasing
$\rho_0$ the experimentally observed $I(V)$ hysteresis disappears
more quickly than expected from theory. One could speculate that
the larger $\rho_0$ the more strongly degraded is the contact
structure, simultaneously broadening the steep rise of $\rho (T)$
around the AFM transition, see inset in Fig.~1 in
Ref.~\cite{Kvitnitskaya}. Such a broadening would be similar to
that of the SC transition of the PC in Fig.~\ref{rt}.

The slightly reduced size of the hysteresis loops in an applied
magnetic field, shown in the lower inset of Fig.~\ref{iv1}, goes
in the same direction. This could be attributed to a small
positive magnetoresistivity of UPd$_2$Al$_3$ \cite{Sugawara}.

Note that the AFM transition itself is difficult to resolve in the
$I(V)$ characteristics. This transition shows up as a small step
in the derivative of $\rho (T)$. Since $I(V)$ is described by an
integral containing $\rho (T)$ over a certain range of
temperatures defined by the bias voltage, one would at least have
to check the second derivative $d^2I/dV^2$. Nevertheless, the huge
anomalies in $I(V)$ reflect the AFM transition because the
magnetic ordering improves dramatically the coherence of the
electron scattering processes, leading to the steeply decreasing
resistivity.

In the $I(V)$ characteristics SC appears as an 'excess' current,
see the upper inset of Fig.~\ref{iv1}. To calculate $I(V)$ of  the
superconducting anomaly we assumed that $\rho (T)$ varies like the
contact resistance $R(T)$ (see Fig.\,\ref{rt}), normalized to  the
normal-state $\rho_0$. This lead to a single peak at around
0.15\,mV, while the experimental $I(V)$ in the upper inset of
Fig.~\ref{iv1} rises almost continuously. Thus the thermal model
\cite{Verkin,Kulik}, developed for normal-state contacts, fails to
describe even qualitatively the resistance of the SC contacts.
This failure could have two reasons. First, the broad SC
transition indicates that the contact does not have a single
$T_c$, but a whole distribution ranging from $T_c \approx 0$ at
the center of the contact where it is reduced due to stress and
disorder and $T_c = 1.8\,$K far away in the indisturbed bulk
material.

A multiply connected contact, where each connection has its own
$T_c$ creating a single spike in the spectrum, could be excluded
because Sharvin's resistance was recovered. Consequently also the
normal-state residual resistivity may vary locally, both along and
vertical to the contact direction, while Eqs.~(\ref{T-V}) and
(\ref{IVT}) have been derived for homogeneous samples only. This
would greatly affect the $I(V)$ characteristic at low bias
voltages since the pattern of current flow could change abruptly,
for example when the critical supercurrent is exceeded in part of
the contact region. It will not change $I(V)$ at large bias
voltages because then the large $T$-dependent part of the
electrical resistivity takes over. Second, at low temperatures,
when both the elastic and the ineleastic electron mean free paths
are largest, the UPd$_2$Al$_3$ contacts could be in the diffusive
instead of the thermal regime. With increasing temperature or bias
voltage the mean free paths get smaller, and the contact is forced
into the thermal regime again.

Break junctions with URu$_2$Si$_2$, another heavy-fermion SC, had
peaks in the differential resistance $dV/dI(V)$ at voltages
described by Eq.~(\ref{T-V}) \cite{Naidyuk1}. This indicated the
destruction of SC in the constriction due to local heating. We
have observed the same behavior also for our UPd$_2$Al$_3$
contacts in the SC state.

The UPd$_2$Al$_3$ junctions presented here are non-linear devices.
Their $N$-shaped $I(V)$ characteristics have a negative
differential resistance. Those devices could be applied -- in
principle -- like an Esaki tunnel diode or a Gunn diode as
amplifiers,  generators, or switching units \cite{Esaki,Price}. Of
practical interest is therefore the possible minimum response
time. We estimate it by the thermal relaxation time $\tau \simeq
(c D/ \lambda)d^2$ of the contact \cite{Verkin}. Here $c$ is the
thermal heat capacity per volume, $D$ is material density and
$\lambda$ the thermal conductivity. With the molar heat capacity
of $3.5\,$J/K$^{-1}$mol$^{-1}$ \cite{Geibel}, $D\approx
10\,$g\,cm$^{-3}$, and $\lambda \approx 4\,$W\,K$^{-1}$m$^{-1}$
\cite{Hiroi}  at 10\,K the relaxation time becomes $\tau \approx
100\,$ps for a $d = 100\,$nm wide contact. This is three orders of
magnitude larger than for a standard tunnel diode, but it could be
reduced by using smaller contacts as long as they remain in the
thermal regime. One (dis)advantage, however, is the low $4\,$mV
working point (at the maximum negative slope of $I(V)$), an order
of magnitude below that of typical Esaki tunnel diodes.

\section{Conclusion}

Sub-$\mu$m scale metallic break-junctions of heavy-fermion
UPd$_2$Al$_3$ showed hysteretic $I(V)$ characteristics at low
temperatures. These highly nonlinear $I(V)$ curves can be
reproduced theoretically by assuming the constrictions are in the
thermal regime. Such anomalous $I(V)$ curves are due to the
unusual $\rho(T)$ dependence of UPd$_2$Al$_3$ at the AFM
transition. Since those point contacts with $N$-shaped $I(V)$
characteristics are nonlinear elements with a negative
differential resistance, they could serve as the analogue of Esaki
tunnel diodes or Gunn diodes as amplifiers, generators, and
switching units. From this point of view UPd$_2$Al$_3$ is not such
a unique material -- each metal with similar $\rho (T)$ should
produce similar $I(V)$ characteristics. This can be expected for
many materials which order magnetically at low temperatures, since
their resistivity typically increases steeply when magnetic order
is destroyed by thermal fluctuations.

{}

\end{document}